\documentclass{aa}  

\usepackage{graphicx}
\usepackage{txfonts}
\begin{document}

   \title{The origin of X-ray emission in 3CRR sources: Hints from mid-infrared Spitzer observations}

   \titlerunning{Origin of X-ray emission in 3CRR sources}

   \author{Shuang-Liang Li
          \inst{1,2}
          \and
          Minfeng Gu
          \inst{1}
          }
   \authorrunning{S.-L. Li \& M. Gu }
   
   \institute{Key Laboratory for Research in Galaxies and Cosmology, Shanghai Astronomical Observatory, Chinese Academy of Sciences, 80 Nandan Road, Shanghai 200030, China\\
              \email{lisl@shao.ac.cn}
         \and
             University of Chinese Academy of Sciences, 19A Yuquan Road, 100049, Beijing, China\\
             }

   \date{Received; accepted}

% \abstract{}{}{}{}{} 
% 5 {} token are mandatory
 
  \abstract
   {}
   {Whether X-ray emission in  radio-loud (RL) active galactic nuclei (AGNs) originates from disk coronae or jets is still under debate. For example, the positive relationships in radio-quiet (RQ) AGNs (such as the optical to X-ray spectral index $\alpha_{\rm {OX}}$ and Eddington ration $\lambda_{\rm {O}}$ as well as the X-ray photon index $\Gamma$ and $\lambda_{\rm {O}}$) are not detected in RLAGNs. We intend to further investigate this issue in this work.}
   {A sample of 27 luminous sources (including 16 quasars and 11 high-excitation radio galaxies) was selected from the 3CRR catalog to reinvestigate the origin of X-ray emission in RLAGNs, where the X-ray and mid-infrared fluxes are observed by Chandra/XMM-Newton and Spitzer, respectively.}
   {It is found for the first time that there is a significant relationship between the mid-infrared to X-ray spectral index $\alpha_{\rm {IX}}$ and $\lambda_{\rm {IR}}$ for whole sample, while there is no relationship between $\alpha_{\rm {OX}}$ and $\lambda_{\rm {O}}$ in quasars. There are strong positive relationships between both $L_{\rm {R}}$-$L_{\rm {X}}$ and $L_{\rm {UV}}$-$L_{\rm {X}}$ panels, which can be well fitted by the disk-corona model. However, there is no significant relationship between $\Gamma$ and $\lambda_{\rm {IR}}$. The possible reason is related to the effects of the large-scale magnetic field in RLAGNs.}
   {We suggest that the X-ray emission in high-excitation RLAGNs originates from a disk-corona system.}


   \maketitle
%
%-------------------------------------------------------------------

\section{Introduction}

The gravitational energy released in the accretion process is believed to be the energy reservoir of active galactic nuclei (AGNs), while the origin of X-ray emission is still a debatable issue. For low-luminosity AGNs, whose Eddington ratio is usually lower than $\sim 10^{-2}$ \citep{2008ARA&A..46..475H}, a strong relationship between radio luminosity $L_{\rm R}$ and X-ray luminosity $L_{\rm X}$ ($L_{\rm R}\propto L_{\rm X}^{0.7}$) has been reported by several works (e.g., \citealt{2003MNRAS.345.1057M,2004A&A...414..895F}). This can be understood well by the accretion-jet model \citep{2014ARA&A..52..529Y}, where the radio and X-ray emissions are considered to be from jets and accretion flows, respectively. However, a steeper correlation ($L_{\rm R}\propto L_{\rm X}^{1.4}$) is given by some authors in FRI (defined by edge-darkened radio lobes, \citealt{1974MNRAS.167P..31F}) radio galaxies (e.g., \citealt{2015MNRAS.453.3447D}), where the X-ray emission is regarded as originating from jets. The multiwavelength emission of FRI radio galaxies has been extensively investigated in recent years \citep{2007A&A...471..137C, 2007ApJ...669...96W, 2009MNRAS.396.1929H, 2009ApJ...701..891L,2010ApJ...725.2426B, 2012A&AT...27..557A}. While the emission at radio and optical bands is believed to be from jets, the origin of infrared and X-ray emission is still controversial (see \citealt{2012A&AT...27..557A} for a review). For example, \citet{2009MNRAS.396.1929H} stated that both nonthermal synchrotron and thermal dust emission can be responsible for the mid-infrared radiation in FRI radio galaxies. The X-ray emission is suggested to originate from a radiation-inefficient accretion flow (RIAF, \citealt{2014ARA&A..52..529Y}) or a jet at a relatively high ($\geq 10^{-4}$) or low ($\leq 10^{-6}$) Eddington ratio, respectively \citep{2007ApJ...669...96W}. FRI radio galaxies are mainly composed of low-excitation radio galaxies, but they may contain a small fraction of high-excitation radio galaxies as well, whose central accretion flows are believed to be different \citep{2007MNRAS.376.1849H, 2009MNRAS.396.1929H,2012MNRAS.421.1569B}. \citet{2018MNRAS.481L..45L} presented a shallower slope in the $L_{\rm R}-L_{\rm X}$ relationship ($L_{\rm R}\propto L_{\rm X}^{0.63}$) by reinvestigating only the low-excitation radio galaxies, indicating an accretion flow origin of X-ray therein.

The situation seems to be more complicated in luminous AGNs, which can be divided into radio-quiet (RQ) and radio-loud (RL) AGNs according to their radio loudness $R$ ($R=f_{\rm {5GHz}}/f_{\rm {4400{\AA}}}$, where $f_{\rm {5GHz}}$ and $f_{\rm {4400{\AA}}}$ are the radio flux at 5GHz and optical flux at 4400${\rm\AA}$, respectively, \citealt{1989AJ.....98.1195K}). The accretion process can be well described by a disk-corona model in RQAGNs \citep{2012MNRAS.422.3268J,2017A&A...602A..79L,2018MNRAS.477..210Q}, where the optical-UV flux is emitted from an optically thick, geometrically thin accretion disk and the hard X-ray flux comes from the inverse Compton scattering of optical-UV photons by the hot corona. This model can naturally fit the observational positive relationships between $\alpha_{\rm OX}$ and $\lambda_{\rm O}$ as well as $\Gamma$ and $\lambda_{\rm O}$ in RQAGNs \citep{2006ApJ...644...86S,2009ApJ...700L...6R,2010A&A...512A..34L,2015A&ARv..23....1B,2019MNRAS.490.3793L}, where the optical to X-ray spectral index $\alpha_{\rm {OX}}$ is defined as $\alpha_{\rm {OX}} =0.384 \log\left[L_\nu {\rm(2500\ \AA)}/L_\nu {\rm(2\ keV)}\right]$ (e.g., \citealt{2010A&A...512A..34L}), and $\lambda_{\rm {O}}=8.1 \nu L_{\nu} (5100 {\rm \AA}) / \textit{L}_{\rm {Edd}}$ is the Eddington ratio based on the optical luminosity at 5100 $\AA$ \citep{2012MNRAS.422..478R}, with $L_{\rm {Edd}}$ being the Eddington luminosity. In contrast, the origin of X-ray emission is still unclear in RLAGNs. In observations, RLAGNs appear to have different properties than RQAGNs, for instance the positive relationships listed above in RQAGNs have not been found in RLAGNs \citep{2019MNRAS.490.3793L,2020ApJ...893...39Zhou}. Furthermore, the average X-ray luminosity in RLAGNs is found to be 2-3 times higher than that in RQAGNs \citep{1981ApJ...245..357Z,1987ApJ...323..243W,2013ApJ...763..109W,2018MNRAS.480.2861G}. This seems to indicate that the X-ray emission from jets is important in RLAGNs. However, \citet{2018MNRAS.480.2861G} recently compiled an excellent sample to investigate the differences of X-ray properties  between luminous radio galaxies and their radio-quiet counterparts, where the black hole mass and Eddington ratio were well selected and the bolometric luminosity were calculated from mid-infrared emission observed by the Wide-field Infrared Survey Explorer (WISE) mission. They argue that the X-ray emission in radio-loud radio galaxies should also come from a disk-corona system because their distribution of the X-ray slope is very similar with those of radio-quiet counterparts (see \citealt{2018MNRAS.480.2861G} for details, see also \citealt{2020MNRAS.492..315G}).

In order to further explore the origin of X-ray emission in RLAGNs, we reinvestigated the $\alpha_{\rm IX}$-$\lambda_{\rm IR}$ and $\Gamma$ -$\lambda_{\rm IR}$ relationships by using the 3CRR catalog through mid-infrared observations at 15 ${\rm {\mu m}}$ from Spitzer, where $\alpha_{\rm IX}$ is the infrared to X-ray spectral index (see equation \ref{ixde} for details) and $\lambda_{\rm {IR}}=8.51 \nu L_{\nu} (15\mu \rm {m}) / \textit{L}_{\rm {Edd}}$ is the Eddington ratio based on the mid-infrared luminosity at $15\mu \rm {m}$ \citep{2012MNRAS.426.2677R}. The reasons we adopted mid-infrared instead of optical-UV emission are as follows. 1) We wanted to take advantage of the bolometric luminosity calculated from mid-infrared emission and compare that with the results of \citet{2018MNRAS.480.2861G}. 2) The big blue bump is obscured in a fraction of reddened quasars (see, e.g., \citealt{2011ApJS..196....2S,2020ApJ...893...39Zhou}). For these sources, it would be better to estimate the bolometric luminosity through infrared emission because optical-UV emissions cannot provide an accurate estimate for the bolometric luminosity. 3) Generally, the mid-infrared emission from torus is more isotropic compared with optical-UV emission from the accretion disk (e.g., \citealt{2006ApJS..166..470R,2015MNRAS.449.1422M}).

\section{The luminous radio galaxy sample}\label{sample}

\begin{figure}
\centering
\resizebox{\hsize}{!}{\includegraphics{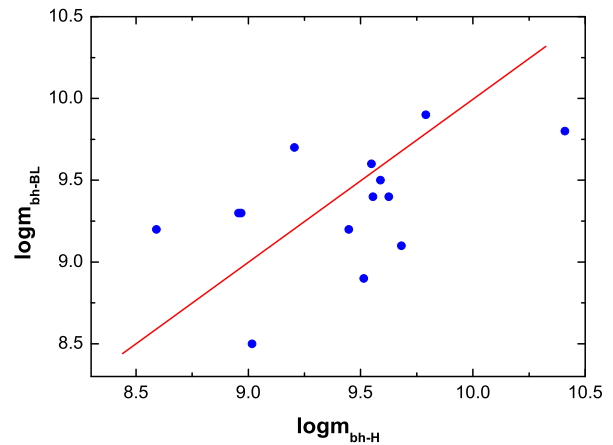}}
\caption{Black hole mass of 15 quasars, where the vertical and horizontal axes represent the black hole mass calculated with different methods. The red line corresponds to  $m_{\rm bh-BL}(=M_{\rm bh-BL}/M_{\odot})=m_{\rm bh-BL}(=M_{\rm bh-H}/M_{\odot})$, where $M_{\rm bh-BL}$, $M_{\rm bh-H}$, and $M_{\odot}$ are the black hole mass calculated with a broad line, the black hole mass calculated with the H band luminosity of the host bulge, and the solar mass, respectively. The average error is about 0.5 dex.  }  \label{mass}
\end{figure}

We intended to study the origin of X-ray emission in luminous radio-loud sources by utilizing the mid-infrared observations at 15 ${\rm {\mu m}}$ from Spitzer. Here "luminous" refers to both high bolometric luminosity ($L_{\rm bol} \geq 3.4\times 10^{44} \rm erg s^{-1}$) and the accretion flow being dominated by an optically thick, geometrically thin disk. We required that the AGNs should be luminous because the accretion flows for luminous and low-luminosity AGNs are different with a critical Eddington ratio $\sim$ 0.01 (e.g., \citealt{2006MNRAS.370.1893H,2014ARA&A..52..529Y}). Most of the objects in our sample have an Eddington ratio higher than 0.01, indicating their accretion flow should be an optically thick, geometrically thin disk. For the other three sources with an Eddington ratio lower than 0.01, their excitation indexes ($\rm EI = \log(O_{\rm III}/H_{\rm \beta})-1/3[log(N_{\rm II}/H_{\rm \alpha})+ log(S_{\rm II}/H_{\rm \alpha}) + log(O_{\rm I}/H_{\rm \alpha})]$, see \citealt{2010A&A...509A...6B} for details) are all higher than 0.95, indicating that they are all high-excitation radio galaxies with an optically thick and geometrically thin disk \citep{2006MNRAS.370.1893H,2009MNRAS.396.1929H, 2010A&A...509A...6B, 2012MNRAS.421.1569B}. After checking \citet{2010A&A...509A...6B} and \citet{2016RAA....16..136H}, we found that all of the radio galaxies in our sample are high-excitation sources.

Starting from the 3CRR catalog\footnote{https://3crr.extragalactic.info/cgi/database}, we firstly selected the objects with an available black hole mass estimation, thus the Eddington ratio, which is a key parameter to investigate X-ray properties, can be calculated (e.g., \citealt{2015A&ARv..23....1B,2018MNRAS.480.2861G,2019MNRAS.490.3793L}). The black hole masses in 84 out of the total 173 objects were measured, where all of them were taken from \citet{2006MNRAS.368.1395M} or \citet{2010A&A...509A...6B}. The mid-infrared emission of 3CRR sources have been presented in several works with Spitzer (e.g., \citealt{2006ApJ...647..161O,2007ApJ...660..117C}). In order to include as many sources as possible, we searched Spitzer IRS spectroscopic data from CASSIS\footnote{http://cassis.astro.cornell.edu} and found that 65 of the 84 objects had been observed by Spitzer. Rest frame 15 $\rm \mu m$ is covered by IRS spectra in all of our sources, thus the flux density at 15 $\rm \mu m$ in the source rest frame can be readily obtained.

Numerous works and telescopes have been devoted to observe the X-ray emissions of 3CRR sources (e.g., \citealt{1999MNRAS.309..969H,2006MNRAS.366..339B,2010ApJ...714..589M,2013ApJ...773...15W}). We required that the objects were observed with X-ray telescopes at a high spatial resolution, with a half-power diameter (HPD) less than $15''$ \citep{2010SPIE.7803E..0HO}. As a result, 14 objects were excluded since the high-resolution data were not found. In the remaining 51 objects (all observed by Chandra or XMM-Newton), the X-ray fluxes were taken from either NED\footnote{ http://ned.ipac.caltech.edu/} or \citet{2020ApJ...893...39Zhou}. The latter compiled an X-ray sample including all 3CRR quasars observed by Chandra. Among these objects, we further ruled out nine FRI radio galaxies, whose accretion process is usually radiatively inefficient (e.g., \citealt{2007MNRAS.376.1849H}); 11 compact steep spectrum (CSS) radio sources, whose central engine is still not fully understood; and three objects whose radio-loudness is lower than 10. The well-known blazar 3C 454.3 was also excluded. Finally, we were left with a sample of 27 luminous radio-loud objects, including 11 high-excitation radio galaxies and 16 quasars (see table 1). None of the objects in our sample are blazars.

Columns (1), (2), and (4) in table 1 were taken from the 3CRR catalog$^1$. The black hole masses in column (3) were all taken from \citet{2006MNRAS.368.1395M} and \citet{2010A&A...509A...6B}. The black hole masses were calculated with different methods in our sample, that is, the width of broad lines for quasars and the H band luminosity of the bulge for radio galaxies. In order to estimate the differences of these two methods, we searched NED for the bulge luminosity in the quasar sample and get the mass of 15 quasars (except 3C 181 and 3C275.1), which were observed at the H band (see column 5 in table 2). The black hole masses according to the two methods are given in Figure \ref{mass}, which are consistent within the errors considering the typical 0.5 dex uncertainty of broad-line-based black hole mass \citep{2000ApJ...533..631Kaspi}. We collected the mid-infrared spectral luminosity $L_{\rm IR}$ at 15 $\mu \rm {m}$ in column (5) through CASSIS. For objects without direct observations at rest frame $15 {\rm {\mu m}}$, we derived the flux at $15 {\rm {\mu m}}$ by the closest wavelength with a slope index of $\alpha_{\rm {IR}}=-1$ ($f_{\nu}\sim \nu^{-\alpha_{\rm {IR}}}$, e.g., \citealt{2011ApJS..196....2S}). The error bars for each piece of data were also included if they were available. For the mid-infrared luminosity taken from Spitzer, the flux uncertainty is less than 10\% (e.g., \citealt{2006ApJ...647..161O}).

In order to compare this with previous results in RQ quasars, we searched the optical-UV data from the literature for the 17 quasars in our sample. The  optical-UV data given in table 2 were observed by the Sloan Digital Sky Survey (SDSS, \citealt{2020ApJ...893...39Zhou}).

\begin{table*}
\normalsize
%\flushleft
\centering
\caption{3CRR sample.}
%\begin{minipage}{\textwidth}
%\resizebox{\textwidth}{!}
%\centering
%\begin{center}
\begin{tabular}{llllllrllll}
\hline
{Name} & {$z$} & {log($\frac{M_{\rm BH}}{M_{\odot}}$)} &  {log$L_{\rm R}$} &
{log$L_{\rm IR}$} &  {log$L_{\rm X}$} &
{log$\lambda_{\rm {IR}}$} &  {log$R_{\rm UV}$} &  {$\Gamma$}  &  {$\alpha_{\rm IX}$} & {Type}\\
{} &  {} &  {} &  {[$\rm erg s^{-1} Hz^{-1}$]} &
 {[$\rm erg s^{-1} Hz^{-1}$]} &  {[$\rm erg s^{-1} Hz^{-1}$]} &  {} &  {}  &  {} & {}  & {}\\
{(1)} &  {(2)} &  {(3)} &  {(4)} &
 {(5)} &  {(6)} &  {(7)} &  {(8)}  &  {(9)} & {(10)}  & {(11)}\\
\hline

3C 33    &  0.0595 &  8.7$^{(1)}$  &  30.26  &    30.68  &  25.86 & -1.90  &    1.12 &   1.7$^{(a)}$  &    1.10  &  HEG\\
3C 33.1  &  0.181 &  8.6$^{(1)}$  &  31.02  &    31.46  &  25.89$^{+0.01}_{-0.01}$ & -1.03  &    1.11 &   1.2$^{(b)}$  &    1.27$^{+0.004}_{-0.004}$  &  BLRG\\
3C 47    &  0.425 &  9.2$^{(2)}$  &  32.65  &    32.09  &  27.29$^{+0.25}_{-0.17}$ & -0.99  &    2.10 &   1.87$^{+0.21(c)}_{-0.22}$ &    1.10$^{+0.04}_{-0.06}$  &  BLQ\\
3C 61.1  &  0.186 &  8.1$^{(1)}$  &  30.35  &    30.27  &  24.96$^{+0.07}_{-0.07}$ & -1.71  &    1.62 &   2.0$^{+0.15(b)}_{-0.15}$  &    1.21$^{+0.006}_{-0.006}$  &  HEG \\
3C 79    &  0.256 &  9.0$^{(1)}$  &  31.27  &    31.80  &  26.79$^{+0.57}_{-0.53}$ & -1.09  &    1.02 &   1.77$^{+0.69(d)}_{-0.61}$ &    1.14$^{+0.11}_{-0.13}$  &  HEG \\
3C 109   &  0.306 &  8.3$^{(2)}$  &  32.87  &    32.42  &  27.34 & 0.24   &    1.99 &   1.69$^{(e)}$ &    1.16  &  BLRG \\
3C 175   &  0.768 &  9.9$^{(2)}$  &  32.79  &    32.42  &  27.50$^{+0.08}_{-0.07}$ & -1.37  &    1.91 &   1.52$^{+0.40(c)}_{-0.07}$ &    1.12$^{+0.02}_{-0.02}$  &  BLQ\\
3C 181   &  1.382 &  9.6$^{(2)}$  &  32.83  &    32.60  &  27.54$^{+0.10}_{-0.08}$ & -0.89  &    1.78 &   1.72$^{+0.17(c)}_{-0.10}$ &    1.15$^{+0.02}_{-0.02}$  &  BLQ \\
3C 196   &  0.871 &  9.6$^{(2)}$  &  32.39  &    32.63  &  27.21$^{+0.49}_{-0.25}$ & -0.85  &    1.30 &   1.67$^{+0.40(c)}_{-0.38}$ &    1.24$^{+0.05}_{-0.11}$  &  BLQ \\
3C 204   &  1.112 &  9.5$^{(2)}$  &  33.24  &    32.42  &  27.73$^{+0.16}_{-0.14}$ & -0.96  &    2.37 &   2.15$^{+0.14(c)}_{-0.22}$ &    1.07$^{+0.03}_{-0.03}$  &  BLQ \\
3C 207   &  0.684 &  8.5$^{(2)}$  &  33.99  &    32.16  &  27.36$^{+0.19}_{-0.15}$ & -0.23  &    3.38 &   1.59$^{+0.28(c)}_{-0.23}$ &    1.09$^{+0.03}_{-0.04}$  &  BLQ \\
3C 208   &  1.109 &  9.4$^{(2)}$  &  33.52  &    32.26  &  27.43$^{+0.10}_{-0.08}$ & -1.03  &    2.81 &   1.63$^{+0.18(c)}_{-0.14}$ &    1.10$^{+0.02}_{-0.02}$  &  BLQ \\
3C 212   &  1.049 &  9.2$^{(2)}$  &  33.93  &    32.65  &  27.83$^{+0.01}_{-0.01}$ & -0.43  &    2.82 &   1.85$^{+0.03(c)}_{-0.08}$ &    1.10$^{+0.003}_{-0.003}$  &  BLQ \\
3C 219   &  0.174 &  9.2$^{(1)}$  &  31.61  &    30.89  &  26.03 & -2.19  &    2.26 &   1.58$^{(e)}$ &    1.11  &  BLRG \\
3C 234   &  0.185 &  9.3$^{(1)}$  &  31.91  &    32.21  &  27.02 & -0.97  &    1.24 &   1.53$^{+0.03(f)}_{-0.04}$ &    1.18  &  HEG \\
3C 245   &  1.029 &  9.4$^{(2)}$  &  34.69  &    32.89  &  27.61$^{+0.04}_{-0.03}$ & -0.39  &    3.34 &   1.57$^{+0.10(c)}_{-0.05}$ &    1.20$^{+0.01}_{-0.01}$  &  BLQ \\
3C 249.1 &  0.311 &  9.3$^{(2)}$  &  32.32  &    31.92  &  26.82$^{+0.08}_{-0.08}$ & -1.26  &    1.94 &   1.92$^{+0.13(c)}_{-0.04}$ &    1.16$^{+0.02}_{-0.02}$  &  BLQ \\
3C 254   &  0.734 &  9.3$^{(2)}$  &  32.65  &    32.25  &  27.21$^{+0.07}_{-0.08}$ & -0.93  &    1.94 &   1.64$^{+0.11(g)}_{-0.10}$ &    1.15$^{+0.02}_{-0.02}$  &  BLQ \\
3C 263   &  0.652 &  9.1$^{(2)}$  &  33.42  &    32.51  &  27.22$^{+0.07}_{-0.06}$ & -0.47  &    2.46 &   1.88$^{+0.10(g)}_{-0.10}$ &    1.21$^{+0.01}_{-0.02}$  &  BLQ \\
3C 268.4 &  1.402 &  9.8$^{(2)}$  &  33.76  &    32.97  &  27.61$^{+0.11}_{-0.09}$ & -0.71  &    2.33 &   1.45$^{+0.17(c)}_{-0.12}$ &    1.22$^{+0.02}_{-0.03}$  &  BLQ \\
3C 275.1 &  0.557 &  8.3$^{(2)}$  &  33.18  &    31.77  &  26.31$^{+0.03}_{-0.03}$ & -0.41  &    2.95 &   1.45$^{+0.05(g)}_{-0.05}$ &    1.24$^{+0.003}_{-0.006}$  &  BLQ \\
3C 300   &  0.272 &  8.8$^{(1)}$  &  31.28  &    30.30  &  25.76 & -2.39  &    2.53 &   1.78$^{+0.06(d)}_{-0.06}$ &    1.03  &  HEG \\
3C 334   &  0.555 &  9.7$^{(2)}$  &  33.11  &    32.41  &  27.09$^{+0.04}_{-0.04}$ & -1.18  &    2.25 &   1.85$^{+0.18(c)}_{-0.11}$ &    1.21$^{+0.01}_{-0.01}$  &  BLQ \\
3C 380   &  0.691 &  8.9$^{(2)}$  &  35.17  &    32.58  &  27.65$^{+0.08}_{-0.08}$ & -0.20  &    4.18 &   1.54$^{+0.09(g)}_{-0.09}$ &    1.13$^{+0.02}_{-0.02}$  &  BLQ \\
3C 382   &  0.0578&  9.3$^{(1)}$  &  31.13  &    30.92  &  26.47$^{+0.002}_{-0.002}$ & -2.26  &    1.76 &   1.80$^{(h)}$ &    1.01$^{+0.00}_{-0.00}$  &  BLRG \\
3C 390.3 &  0.0569&  8.8$^{(1)}$  &  31.35  &    31.00  &  26.49 & -1.68  &    1.90 &   1.85$^{+0.02(i)}_{-0.02}$ &    1.03  &  BLRG \\
3C 452   &  0.0811&  8.8$^{(1)}$  &  31.28  &    30.78  &  26.40 & -1.91  &    2.04 &   1.70$^{(j)}$ &    1.00  &  HEG \\

\hline
\end{tabular}\\
%\end{center}
%\end{minipage}
Notes: Col. (1): Source name. Col. (2): Redshift. Col. (3): Black hole mass. Col. (4): Radio spectral luminosity at 5 GHz. Col. (5): Mid-infrared spectral luminosity at 15 $\rm {\mu m}$. Col. (6): X-ray spectral luminosity at 2 keV. Col. (7): Eddington ratio $\lambda_{\rm {IR}}$ based on mid-infrared flux. Col. (8): Radio loudness $R_{\rm UV}\equiv L_{\rm R}/L_{\rm {4400{\AA}}}$. Col. (9): Photon index $\Gamma$ at 2-10 kev ($f_\nu \sim \nu^{1-\Gamma}$). Col. (10): Mid-infrared to X-ray spectral index $\alpha_{\rm IX}\equiv 0.228 \log\left[L_{\rm {IR}}/L_{\rm {X}}\right]$. Col. (11): Type of objects. Here HEG, BLRG, and BLQ are the acronyms used for high-excitation radio galaxies, broad-line radio galaxies, and broad-line quasars, respectively \citep{2010A&A...509A...6B}.\\
$^{(1)-(2)}$: References for black hole mass. (1): \citet{2010A&A...509A...6B}; (2): \citet{2006MNRAS.368.1395M}.
\\
$^{(a)-(g)}$: References for photon index. (a): \citet{2009A&A...498...61T}; (b): \citet{2010ApJ...714..589M}; (c): \citet{2020ApJ...893...39Zhou}; (d): \citet{2006MNRAS.370.1893H}; (e): \citet{2005ApJ...629...88S}; (f): \citet{2008A&A...480..671P}; (g): \citet{2006MNRAS.366..339B}; (h):\citet{2011ApJ...734..105S}; (i):\citet{2012ApJ...745..107W}; (j):\citet{2006ApJ...642...96E}.
\end{table*}

\begin{table}
\normalsize
%\flushleft
\centering
\caption{3CRR quasar sample .}
%\begin{minipage}{\textwidth}
%\begin{center}
%\centering
\begin{tabular}{lcccc}
\hline
{Name} & {log$L_{\rm {UV}}$} &  {log$\lambda_{\rm O}$} &  {$\alpha_{\rm OX}$}  &  {log($\frac{M_{\rm bh-H}}{M_{\odot}}$)} \\
{} &  {[$\rm erg s^{-1} Hz^{-1}$]} &  {} &  {} &  {} \\
{(1)} &  {(2)} &  {(3)} &  {(4)} &  {(5)} \\
\hline

3C 47    &  30.13 & -1.29  &    1.09$^{+0.01}_{-0.07}$  &  8.59   \\
3C 175   &  31.16 & -0.95  &    1.41$^{+0.03}_{-0.03}$  &  9.79   \\
3C 181   &  31.20 & -0.61  &    1.41$^{+0.04}_{-0.03}$  &     \\
3C 196   &  30.77 & -1.05  &    1.37$^{+0.20}_{-0.10}$  &  9.55   \\
3C 204   &  31.12 & -0.60  &    1.30$^{+0.06}_{-0.05}$  &  9.59   \\
3C 207   &  30.37 & -0.35  &    1.16$^{+0.07}_{-0.06}$  &  9.02   \\
3C 208   &  31.10 & -0.52  &    1.41$^{+0.04}_{-0.03}$  &  9.55   \\
3C 212   &  30.60 & -0.81  &    1.07$^{+0.004}_{-0.004}$  &  9.45   \\
3C 245   &  30.96 & -0.65  &    1.29$^{+0.02}_{-0.01}$  &  9.63   \\
3C 249.1 &  30.50 & -1.02  &    1.41$^{+0.03}_{-0.03}$  &  8.97   \\
3C 254   &  30.72 & -0.79  &    1.35$^{+0.03}_{-0.03}$  &  8.96   \\
3C 263   &  31.20 & -0.12  &    1.53$^{+0.03}_{-0.02}$  &  9.68   \\
3C 268.4 &  31.44 & -0.58  &    1.47$^{+0.04}_{-0.03}$  &  10.4   \\
3C 275.1 &  30.06 & -0.46  &    1.44$^{+0.01}_{-0.01}$  &     \\
3C 334   &  30.59 & -1.33  &    1.34$^{+0.02}_{-0.02}$  &  9.20   \\
3C 380   &  30.95 & -0.17  &    1.27$^{+0.03}_{-0.03}$  &  9.51   \\

\hline
\end{tabular}\\
%\end{center}
%\end{minipage}
Notes: Col. (1): Source name. Col. (2): UV luminosity at 2500$\rm { {\AA}}$. Col. (3): Eddington ratio  $\lambda_{\rm {O}}$ based on optical-UV flux. Col. (4): Optical to X-ray spectral index $\alpha_{\rm {OX}}$. Col. (5): Black hole mass calculated with the luminosity at the H band.
\end{table}

\section{Results}\label{results}

\begin{figure}
\centering
\resizebox{\hsize}{!}{\includegraphics{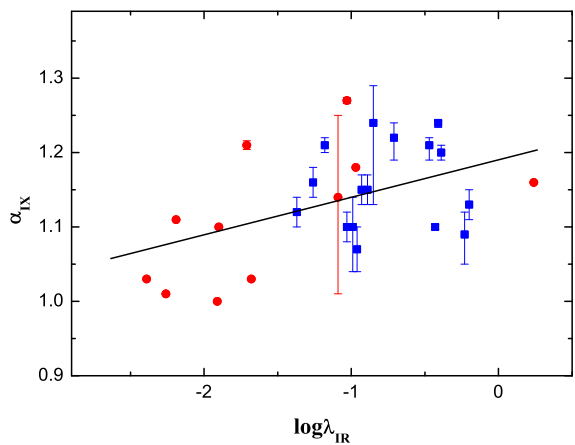}}
\caption{Mid-infrared to X-ray spectral index $\alpha_{\rm {IX}}$ as a function of Eddington ratio $\lambda_{\rm {IR}}$ based on mid-infrared emission for RLAGNs, where the red circle and blue square represent the narrow-line radio galaxies and quasars, respectively. The black solid line shows the best fit given by equation (2).} \label{f1}
\end{figure}

\begin{figure}
\centering
\resizebox{\hsize}{!}{\includegraphics{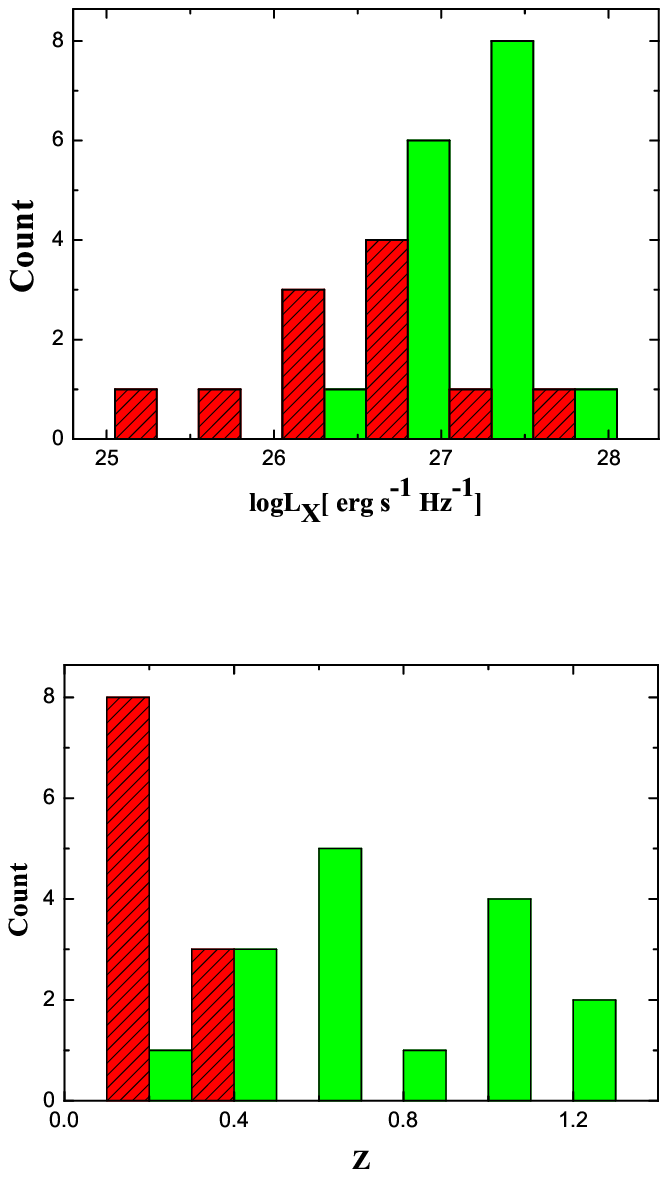}}
\caption{Upper and lower panels: Histograms for X-ray luminosity at 2 kev and redshift, respectively, where the red slash and green regions represent the narrow-line radio galaxies and quasar samples, respectively.} \label{hist1}
\end{figure}

\begin{figure}
\centering
\resizebox{\hsize}{!}{\includegraphics{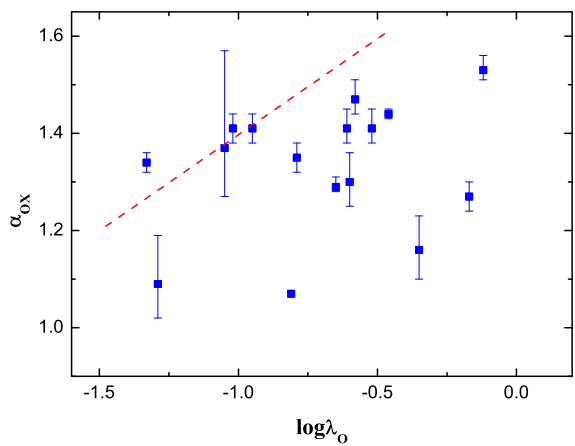}}
\caption{Optical to X-ray spectral index $\alpha_{\rm {OX}}$ as a function of the Eddington ratio $\lambda_{\rm {O}}$ based on optical emission for the quasar sample, where the red dashed line represents the $\alpha_{\rm OX}-\lambda_{\rm O}$ relationship given by \citet{2010A&A...512A..34L}.}  \label{f2}
\end{figure}

\begin{figure}
\centering
\resizebox{\hsize}{!}{\includegraphics{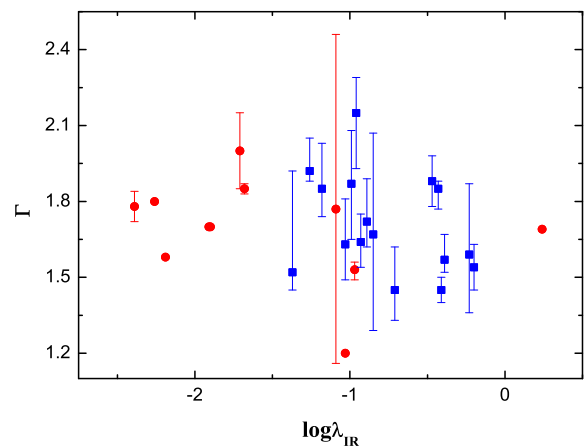}}
\caption{X-ray photon index $\Gamma$ as a function of the Eddington ratio $\lambda_{\rm {IR}}$ based on mid-infrared emission for RLAGNs, where the symbols are the same as in figure \ref{f1}.}  \label{f3}
\end{figure}

\begin{figure}
\centering
\resizebox{\hsize}{!}{\includegraphics{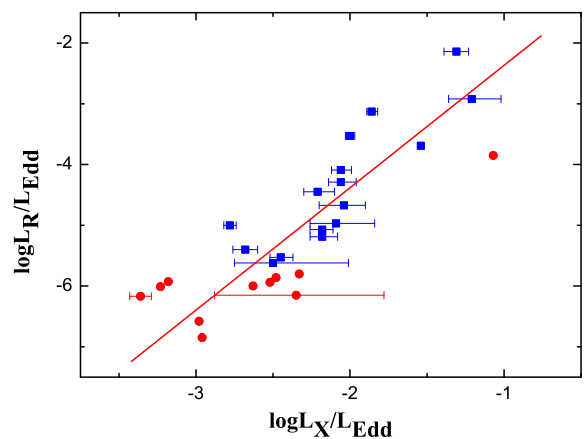}}
\caption{Radio luminosity $L_{\rm {R}}$ as a function of the X-ray luminosity $L_{\rm {X}}$ for RLAGNs, where the symbols are the same as in figure \ref{f1}. The red solid line represents our best fitting result.} \label{f4}
\end{figure}

\begin{figure}
\centering
\resizebox{\hsize}{!}{\includegraphics{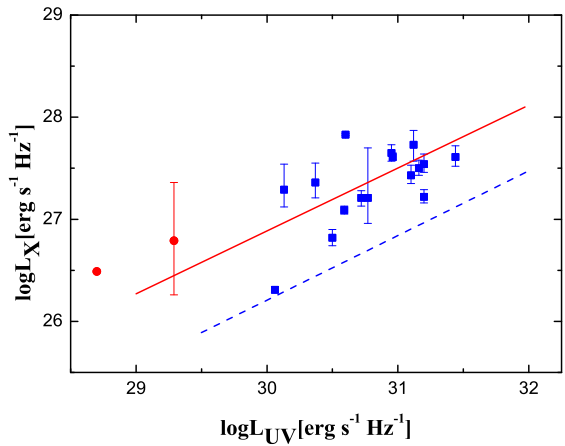}}
\caption{X-ray luminosity $L_{\rm {X}}$ as a function of the optical-UV luminosity $L_{\rm {UV}}$ for RLAGNs, where the symbols are the same as in figure \ref{f1}. The red solid line and blue dashed line represent our best fitting result and the result of RQAGNs given by \citet{2017A&A...602A..79L}, respectively. } \label{f5}
\end{figure}

The average X-ray flux in RLAGNs is found to be higher than that in RQAGNs, while its origin is still unclear (e.g., \citealt{2018MNRAS.480.2861G,2020ApJ...893...39Zhou}). The additional X-ray emission may be attributed to the contribution of either the jet \citep{1987ApJ...323..243W,2019MNRAS.490.3793L,2020ApJ...893...39Zhou} or the fast rotating black holes \citep{2018MNRAS.480.2861G,2020MNRAS.492..315G}. The spinning black hole can improve the radiative efficiency of the accretion disk and jet power. An obvious difference between these two results is the method chosen to calculate the bolometric luminosity: The former adopts optical-UV emission, while the latter adopts mid-infrared emission. 

In order to further explore the different results obtained by using the observational data at different wavebands, we compiled a sample observed by Spitzer from the 3CRR catalog (see table 1). The bolometric luminosity can be calculated with both the optical and mid-infrared emission by using the bolometric corrections given in \citet{2012MNRAS.422..478R,2012MNRAS.426.2677R}. Following the definition of $\alpha_{\rm {OX}}$, we define a new parameter to reflect the ratio of disk to corona emissions, that is, the infrared to X-ray spectral index $\alpha_{\rm {IX}}$ as:
\begin{eqnarray}
\alpha_{\rm {IX}} &=& -\log\left[L_\nu {\rm(15 \mu m)}/L_\nu {\rm(2\ keV)}\right]/ \nonumber\\
& & \log\left[{\nu{\rm(15 \mu m)}}/\nu{\rm(2\ keV)}\right] \nonumber\\
&=& 0.228 \log\left[L_\nu {\rm(15 \mu m)}/L_\nu {\rm(2\ keV)}\right]. \label{ixde}
\end{eqnarray}
The correlation between $\alpha_{\rm {IX}}$ and $\lambda_{\rm {IR}}$ is explored in figure \ref{f1}. In this work, the ordinary least squares (OLS) method \citep{1990ApJ...364..104I} was adopted to fit all the relationships. A moderately strong positive correlation between $\alpha_{\rm {IX}}$ and $\lambda_{\rm {IR}}$ was found, which reads as 
\begin{equation}
\alpha_{\rm {IX}}=(0.05\pm 0.02)\log \lambda_{\rm {IR}}+1.2\pm 0.02 \label{ixlambda}
,\end{equation}
with a confidence level of 99.3\% based on a Pearson test. This positive relationship was found for the first time in RLAGNs and is flatter than the $\alpha_{\rm OX}-\lambda_{\rm O}$ relationship in RQAGNs (e.g., \citealt{2010A&A...512A..34L}), suggesting a disk origin of X-ray emission. In figure \ref{hist1}, we compare the properties of radio galaxies with quasars. It is found that the radio galaxies have a relatively lower luminosity and redshift than the broad-line quasars. We argue that the correlation in figure \ref{f1} cannot be produced by the different methods to estimate $M_{\rm BH}$ in radio galaxies and broad-line quasars since two methods give consistent $M_{\rm BH}$ measurements in broad-line quasars (see figure \ref{mass}). In addition, since $\alpha_{\rm IX}$ is the ratio of infrared luminosity to X-ray luminosity, and Eddington ratio $\lambda_{\rm IR}$ is the ratio of infrared luminosity to black hole mass, the dependence of both the $\alpha_{\rm IX}$ and Eddington ratio $\lambda_{\rm IR}$ on the redshift thus can be eliminated, although the infrared, X-ray luminosity, and black hole mass are all tightly related with redshift. Therefore, the result in Figure \ref{f1} cannot be caused by different redshift distributions in radio galaxies and broad-line quasars.

As a significant correlation between $\alpha_{\rm IX}$ and $\lambda_{\rm IR}$ has been found in our sample, it will be interesting and necessary to check whether our sample also possesses a $\alpha_{\rm {OX}}$-$\lambda_{\rm O}$ relationship, which was lacking from our previous study for RLAGNs \citep{2019MNRAS.490.3793L}. If the relation is present, one would need to investigate why it has not been shown in previous works, for example, maybe partly due to the different sample selection. Otherwise, that may indicate that the mid-infrared emission can be a better indicator than the optical one when calculating the bolometric luminosity, especially in radio galaxies. We find that there is no strong correlation between $\alpha_{\rm {OX}}$ and $\lambda_{\rm O}$ in figure \ref{f2}, where the red dashed line represents the known $\alpha_{\rm {OX}}$-$\lambda_{\rm O}$ relationship for RQ quasars given by \citet{2010A&A...512A..34L}. The $\alpha_{\rm {OX}}$-$\lambda_{\rm O}$ panel for the 17 quasars is inconsistent with that for RQ quasars and most of the objects are systematically lower than the red dashed line. This is because our sample is RL and thus has higher X-ray luminosity compared with that of RQ quasars in \citet{2010A&A...512A..34L}.

The relationship between $\Gamma $ and $\lambda_{\rm {IR}}$ is investigated in figure \ref{f3}. There is no strong relationship between them, which is the same for $\Gamma $ and $\lambda_{\rm O}$ in RLAGNs reported in previous works \citep{2019MNRAS.490.3793L,2020ApJ...893...39Zhou}. The reason may be related to the effects of large-scale magnetic fields (see section \ref{discussion} for details).

The $L_{\rm R}$-$L_{\rm X}$ relationship is also often adopted to study the physical process in AGNs (e.g., \citealt{2003MNRAS.345.1057M}), which is given by
\begin{equation}
\log (L_{\rm R}/L_{\rm{Edd}})=(2.0\pm 0.2)\log (L_{\rm X}/L_{\rm{Edd}})-0.3\pm 0.6 \label{lrlx}
\end{equation}
in figure \ref{f4}, where the confidence level is higher than 99.9\% based on a Pearson test. The slope of the $L_{\rm R}$-$L_{\rm X}$ relationship in RLAGNs is also found to be larger than that in both the low-luminosity AGNs ($\sim$0.7) and RQAGNs ($\sim$1) \citep{2003MNRAS.345.1057M,2008ApJ...688..826L,2020MNRAS.496..245Zhu}. 

We give the $L_{\rm {X}}-L_{\rm UV}$ relation in figure \ref{f5}, which shows a good relationship. In addition to 16 quasars, two radio galaxies detected by HST, that is 3C79 and 3C 390.3, are also included in figure \ref{f5}. The bisector fitting \citep{1990ApJ...364..104I} gives
\begin{equation}
\log L_{\rm {X}}=(0.6\pm0.1)\log L_{\rm UV}+8.4\pm5.0
\end{equation}
with a confidence level higher than 99.9\%, where the blue dashed line was taken from \citet{2017A&A...602A..79L} for RQ quasars. Most of our data are higher than the blue dashed line. However, the slope is quite similar (the slope is $0.63^{+0.02}_{-0.02}$ in \citealt{2017A&A...602A..79L}). Our slope of 0.6 is also found to be consistent with that of RLAGNs recently reported by \citet{2020MNRAS.496..245Zhu}, where the authors suggest a disk-corona origin for the X-ray emission in an RL quasar as well. The X-ray luminosity of RL quasars is about 3 times higher than that of RQ quasars (see figure \ref{f5}), which is roughly consistent with previous results \citep{1981ApJ...245..357Z, 2018MNRAS.480.2861G}.

\section{Discussion} \label{discussion}

In theory, if RLAGNs are powered by the accretion disk-jet system, the flat-spectrum radio emission at 5 GHz has a power-law dependence on the mass loss rate into a jet \citep{2003MNRAS.343L..59H}, which is given by
\begin{equation}
L_{\rm R}/L_{\rm Edd}\sim \dot{m}_{\rm {jet}}^{1.4}\sim \dot{m}^{1.4},\label{lr}
\end{equation}
where $L_{\rm {Edd}}$ is the Eddington luminosity and $\dot{m}$ is the dimensionless mass accretion rate. Next, we assumed that the X-ray emissions in RL quasars originate from corona, resulting in a similar $L_{\rm X}$-$L_{\rm UV}$ relationship for RQ and RL quasars. Therefore, we could adopt the strong correlation between  $L_{\rm X}$ and $L_{\rm UV}$ in a disk-corona model for RQ quasars \citep{2017A&A...602A..79L}, which reads as follows:
\begin{equation}
L_{\rm {X}}/L_{\rm Edd}\sim (L_{\rm {UV}}/L_{\rm Edd})^{4/7}.\label{lx}
\end{equation}
The optical-UV luminosity can be given as 
\begin{equation}
L_{\rm UV}\sim m^{(5-\gamma_o)/4} \dot{m}^{(3+\gamma_{0})/4},   \label{luv} 
\end{equation}
where $\gamma_{0}$ is the index of an optical/UV spectrum and $m=M_{\rm bh}/M_\odot$ is the black hole mass \citep{2017A&A...602A..79L}. For a standard thin disk, $\gamma_{0}$ is equal to $-1/3$. However, the observational optical-UV spectrum in AGNs usually indicates that $\gamma_o$ is  $\sim$ 1 instead of $-1/3$ (see, e.g., \citealt{2008RMxAC..32....1G,2008MNRAS.387L..41L}). Therefore, in order to better compare this with the observational results, we adopted $\gamma_o=1$ in this work. Thus equation (\ref{luv}) can be written as
\begin{equation}
 L_{\rm UV}/ L_{\rm Edd} \sim \dot{m}. \label{luv2}
\end{equation}

Combining equations (\ref{lr}), (\ref{lx}), and (\ref{luv2}), the relationship between $L_{\rm R}$ and $L_{\rm X}$ is
\begin{equation}
L_{\rm {R}}/L_{\rm Edd}\sim (L_{\rm {X}}/L_{\rm Edd})^{2.45}.
\end{equation}
This is consistent with the observational result (equation \ref{lrlx}) in about the 2.5 $\sigma$ level and this can support our assumption on the disk-corona origin of X-ray emission. Therefore, except for the $\Gamma $-$\lambda_{\rm {IR}}$ relationship, all other results can be well fitted with the disk-corona model.
It should be noted that the disk-corona model for luminous radio galaxies cannot be applied to low-excitation radio galaxies, which are thought to be powered by RIAF \citep{2012MNRAS.421.1569B,2014ARA&A..52..529Y}. In the disk-corona model, the optical-UV emission comes from a thin disk extending to the surroundings of a black hole, while it could be mainly from a jet in low-excitation radio galaxies \cite[e.g.,][]{2007A&A...471..137C}. Our results indicate that the X-ray emission is more likely produced in a disk-corona system. However, this does not completely rule out the jet origin of X-ray emission. For example, for the radio-loud quasars with gamma-ray emission, the X-ray emission may come from a jet through the inverse-Compton (IC) upscattering of the cosmic microwave background (CMB, \citealt{2019ApJ...883L...2M}).

The relationship of $\Gamma$  and $\lambda_{\rm {IR}}$ is still absent in RLAGNs (figure \ref{f3}), while the negative $\Gamma$-$\lambda_{\rm {O}}$ relationship is quite strong in RQAGNs \citep{2009MNRAS.399..349G,2015A&ARv..23....1B}, which can be naturally explained by the disk-corona model \citep{2012MNRAS.422.3268J,2017A&A...602A..79L}. Nevertheless, a large-scale magnetic field (e.g., the size of a field line larger than the local disk scale height), which significantly affects the physical properties of disk corona, is believed to play a key role in the formation of the jet in RLAGNs (\citealt{1977MNRAS.179..433B,1982MNRAS.199..883B,2013MNRAS.434.1692B,2013ApJ...765..149C}). For one thing, the photon index of hard X-ray remains almost constant when the magnetic field strength is strong enough \citep{2016ApJ...833...35L}. For another, a large-scale magnetic field can increase the value of the viscosity parameter $\alpha$ \citep{2013ApJ...767...30B,2016MNRAS.457..857S}, which can also decrease the slope of the $\Gamma$-$\lambda_{\rm O}$ relationship \citep{2018MNRAS.477..210Q}. Therefore, we can anticipate that the relationship between $\Gamma$ and $\lambda_{\rm O}$ in RLAGNs will be very weak or even disappear.

\section{Conclusions} \label{conclusion}

Previous studies have indicated that there is no significant relationship between $\alpha_{\rm{OX}}$ and $\lambda_{\rm O}$ in RLAGNs, and they have suggested that the contribution of the jet to the X-ray flux cannot be neglected \citep{2019MNRAS.490.3793L,2020ApJ...893...39Zhou}. However, \citet{2018MNRAS.480.2861G,2020MNRAS.492..315G} compiled a RLAGN sample based on the mid-infrared data observed by WISE (see also \citealt{2020ApJ...901..111K}) and suggested that the X-ray emission comes from the disk corona instead of the jet. We constructed a high-excitation RLAGN sample to further investigate this question. The moderately strong relationship between  $\alpha_{\rm{IX}}$ and $\lambda_{\rm IR}$ (figure \ref{f1}) in this work supports the conclusion of \citet{2018MNRAS.480.2861G,2020MNRAS.492..315G}, though no significant relationships between $\alpha_{\rm{OX}}$ and $\lambda_{\rm O}$ have been reported in the quasar sample. The similar $L_{\rm X}$-$L_{\rm UV}$ relationships between RQ and RL quasars also indicate a disk origin of X-ray emission. Therefore, we suggest that the X-ray emission in high-excitation RLAGNs should originate from a disk-corona system \citep{2012MNRAS.422.3268J,2017A&A...602A..79L}.

\section* {Acknowledgements}

We thank the reviewer for helpful comments. SLL thanks Dr. Fuguo Xie and Erlin Qiao for helpful discussion. This work is supported by the NSFC (grants 11773056, 11873073). This work has made extensive use of the NASA/IPAC Extragalactic Database (NED), which is operated by the Jet Propulsion Laboratory, California Institute of Technology, under contract with the National Aeronautics and Space Administration (NASA). This work also adopts the archival data obtained with the Spitzer Space Telescope, which is operated by the Jet Propulsion Laboratory, California Institute of Technology under a contract with NASA.

\bibliographystyle{aa.bst}
\bibliography{x-ray.bib}

\end{document}